\documentclass[final]{appolb}
\usepackage{graphicx}
\usepackage{amsmath}
\usepackage[colorlinks=true,linkcolor=red, citecolor=blue, urlcolor=blue]{hyperref}

\usepackage{xcolor}

\newcommand{\PbPb}{\ensuremath{\mathrm{Pb+Pb}}}

\newcommand{\pp}{\ensuremath{\mathrm{pp}}}
\newcommand{\alphas}{\ensuremath{\alpha_\mathrm{s}}}
\newcommand{\Martini}{\textsc{martini}}

\newcommand{\Matter}{\textsc{matter}}
\newcommand{\Vishnu}{\textsc{vishnu}}

\newcommand{\Pythia}{\textsc{pythia}}
\newcommand{\FastJet}{\textsc{fastJet}}
\newcommand{\CUJET}{\textsc{cujet}}
\newcommand{\dglv}{\textsc{dglv}}
\newcommand{\amy}{\textsc{amy}}
\newcommand{\JETSCAPE}{\textsc{jetscape}}
\newcommand{\RAA}{\ensuremath{R_\mathrm{AA}}}
\begin{document}
\title{Jet-medium photons as a probe of parton dynamics
\thanks{Talk presented at the XXIX$^{\rm th}$  International Conference on Ultra-relativistic Nucleus- Nucleus Collisions (Quark Matter 2022)}}

\author{Rouzbeh Modarresi Yazdi\thanks{Speaker}$^\mathrm{a}$, Shuzhe Shi$^\mathrm{b}$, Charles Gale$^\mathrm{a}$, Sangyong Jeon$^\mathrm{a}$
\address{$^\mathrm{a}$ Department of Physics, McGill University, 3600 University Street, Montreal, QC, Canada H3A 2T8\\$^\mathrm{b}$Center for Nuclear Theory, Department of Physics and Astronomy, Stony Brook University, Stony Brook, New York 11794–3800, USA}}
\maketitle

\begin{abstract}

Photons resulting from jet-medium interactions offer the opportunity of studying the 
evolving  quark distribution in a heavy ion collision. The spectra of jet-medium 
photons is presented within the \JETSCAPE\, framework for two different energy loss models, \Martini\, and \CUJET. Jet-medium photons can contribute significantly to the spectrum of direct photons in the intermediate $p_T$ range.
\end{abstract}
  
\emph{Introduction} --- 
Photons, real and virtual, are an important probe of the quark gluon plasma. They are produced at all stages of 
the evolution of the plasma and have a mean-free-path larger than the size of the created 
medium. As such, their study and measurement can be a great tool for quantifying 
the properties of QGP as well as jet-energy loss. Here we present a recent calculation of photons produced from jet interactions with the QGP using the \JETSCAPE~\cite{JETSCAPE:2019udz}\, framework. 
To that comprehensive suite, we have added \CUJET~\cite{Xu:2014ica}\, as a low virtuality energy loss module, 
thus allowing for a systematic comparison with \Martini~\cite{Schenke:2009gb}. 
What follows is a brief description of the formalisms, a discussion of jet-medium photons and 
finally the results of the simulation, a discussion, and outlook on future work.

\vspace{3mm}
\emph{Low Virtuality Energy Loss} --- 
\CUJET\, implements the \dglv~\cite{Gyulassy:2000er,Djordjevic:2003zk} inelastic parton splitting rates, computed  
to leading order in the opacity expansion~\cite{Xu:2014ica}. The calculation of the gluon emission rates assumes that 
the hard probe was created at some given time and position inside the plasma and then evolved, scattering from a dynamical medium~\cite{Xu:2014ica} and radiated a gluon. The calculation of \dglv\, rates is performed in the eikonal limit, with the jet 
and the radiated gluon assumed to be distinguishable from the strongly interacting medium 
around them. The Landau--Pomeranchuk--Migdal (LPM) effect manifests itself as a phase factor. Soft, collinear divergences are regulated by a gluon plasmon mass and the light quarks are considered 
massless~\cite{Djordjevic:2003zk}. It is also 
assumed that the radiated gluon does not modify the direction of travel of the incoming parton. In \CUJET\, collisional energy loss occurs via scattering with the QGP medium. This 
is done using the Thoma--Gyulassy model where the HTL gluon propagator includes a natural 
IR regulator~\cite{THOMA1991491}. Finally, \CUJET\, allows for the running of the strong coupling, \alphas , according 
to the one-loop pQCD expression.

\Martini\, implements the AMY-McGill formalism for radiative and elastic scattering 
energy loss on-shell energetic partons in a strongly interacting medium~\cite{Schenke:2009gb}. It solves a rate equation with gain and loss terms for the time-evolving parton distribution. As in \CUJET, radiative energy loss is strictly collinear, with elastic scattering channels providing the momentum broadening via space-like exchanges with the medium. Unlike the \dglv\, model, the original AMY radiative rates, which we use in this work, do not have an explicit time dependence~\cite{Schenke:2009gb}. The reaction rates calculated with AMY are to all orders of the opacity expansion and account fully for the LPM effect. Collisional energy loss is also implemented, again using 
the gluon HTL propagator. Other than gluon radiation and collisional energy loss, \Martini\, also includes $g\to q+\bar{q}$ radiative channel as well as conversion processes $q\to g$ and $g\to q$. The strong coupling in radiative and elastic channels is allowed to run, using the one-loop expression for \alphas.

\vspace{3mm}
\emph{Jet-Medium Photons} --- 
Two dominant mechanisms of jet-medium photon emission are jet bremsstrahlung and 
jet-photon conversion~\cite{Fries:2002kt}. Jet bremsstrahlung has a similar structure as the 
jet gluon bremsstrahlung channel: the jet receives kicks from the medium and emits a photon. 
The differential rate for this process in \Martini\, is given by
\begin{align}
        \frac{\mathrm{d} \Gamma^{\amy}_{q\to q\gamma}}{\mathrm{d} z} (p,z) =\,
	        \frac{e_f^2 \alpha_\mathrm{em} P_{q\to q\gamma}(z)}{[2p\, z(1{-}z)]^2}& [1-f_q((1-z)p)] \nonumber\\
	        &\times\int\,\frac{\mathrm{d}^2 \mathbf{k}_{\perp}}{(2\pi)^2} ~\mathrm{Re} \left[ 2\,\mathbf{k}_{\perp} \cdot \mathbf{g}_{(z,p)}(\mathbf{k}_{\perp}) \right],
    \label{eq.splitting_rate_martini}
\end{align}
where $e_f$ is the fractional charge of quark(anti-quark) and $z\equiv k/p$ is the momentum fraction carried away by the photon. $\mathbf{g}_{(z,p)}(\mathbf{k}_{\perp})$ is the solution to the integral equation\cite{Arnold:2001ba}
    \begin{equation}
        \label{eq:AMY}
            2\,\mathbf{k}_{\perp} 
            =i\, \delta E(z,p,\mathbf{k}_{\perp}) \mathbf{g}(\mathbf{k}_{\perp})+ \int \frac{\mathrm{d}^2\mathbf{q}_{\perp}}{(2\pi)^2}~{C}(\mathbf{q}_\perp) 
            \Big[\mathbf{g}(\mathbf{k}_{\perp}) - \mathbf{g}(\mathbf{k}_{\perp} - z\,\mathbf{q}_{\perp}) \Big]\
    \end{equation}
where $q_{\perp}$ and $k_{\perp}$ are the exchanged and photon transverse momenta, respectively, $C( \mathbf q_\bot)$ is a scattering kernel, and $\delta\,E$ is the energy difference between the initial and final states\cite{Arnold:2001ba}.

Conversion photons arise from soft momentum exchanges of $q (\bar{q})$ jets with 
the QGP medium. At leading order in the strong coupling, the process can go through either 
QCD Compton scattering or $q\bar{q}$ annihilation. Both processes include a Mandelstam-t channel 
where a soft quark is exchanged. A general property of t-channel processes is the preference 
of the outgoing particle to be in the same direction as the incoming owing to $1/t$ term in the matrix 
element and the matrix element is maximum when the relative angle is very nearly zero. As such, we can 
perform an approximation where the entirety of the contribution to the conversion photon rate stems from 
this collinear region and write the rate as
\begin{equation}
    \frac{\mathrm{d}\Gamma^{\gamma\mathrm{-Conv.}}}{\mathrm{d}p\,\mathrm{d}k}(p,k,T) = e^2_f \frac{2\pi\,\alpha\,\alpha_s}{3}\frac{T}{p}\left[\frac{1}{2}\ln{\frac{2pT}{m^2_q} - 0.36149}\right]\,\delta\left(p-k\right).
\end{equation}
where the collinearity is made explicit by the delta function.

    \begin{figure}[!hbt]
        \includegraphics[width=0.75\linewidth]{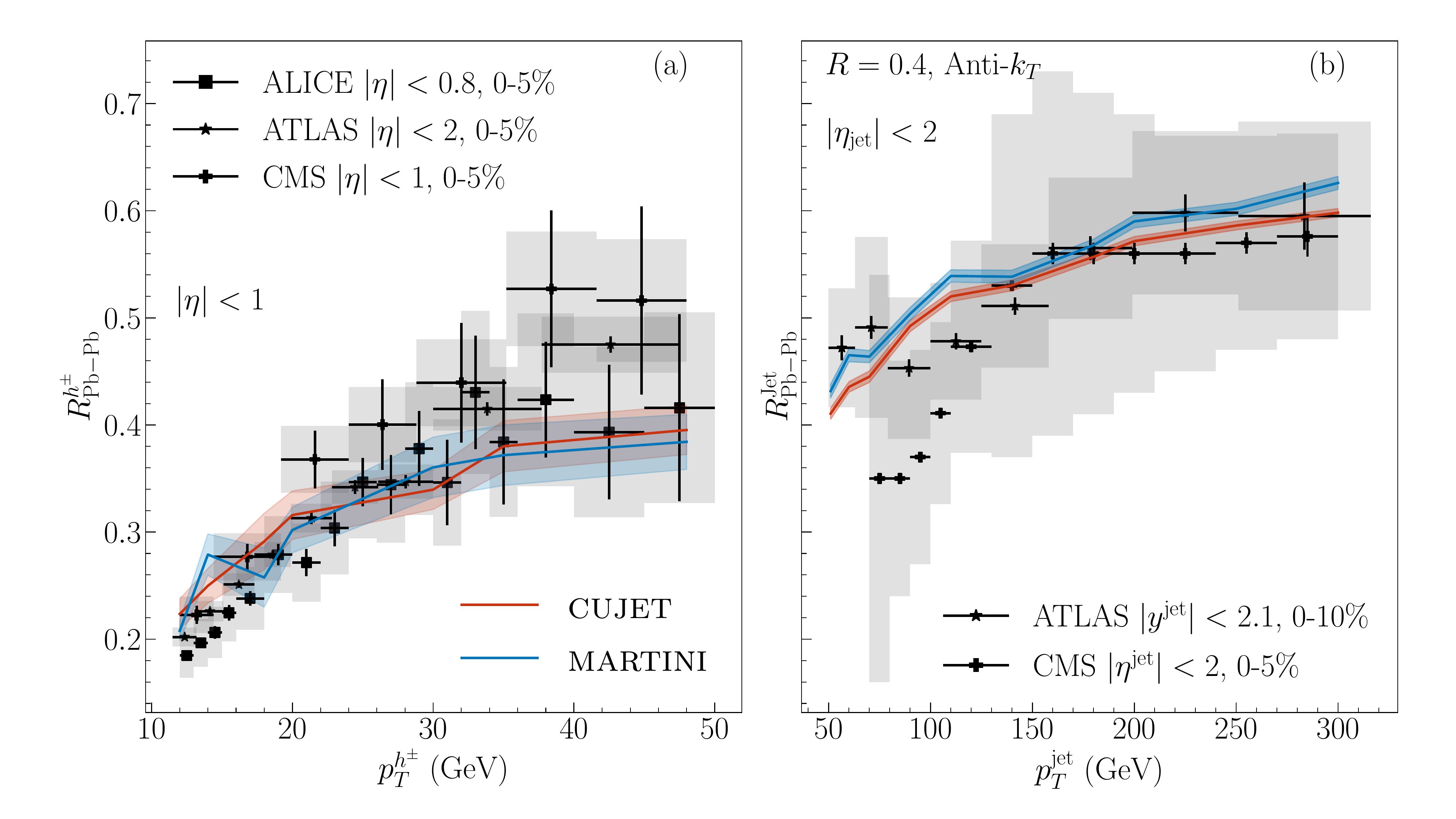}
        \begin{minipage}[b]{7pc}
        \begin{center} 
        \caption{Calculation of (a) charged hadron and (b) jet \RAA\, for \Martini\, and \CUJET\, for \PbPb\,
             collisions at $\sqrt{s}=2.76$ ATeV.
             Data from ~\protect{~\cite{ALICE:2012aqc,ATLAS:2015qmb,CMS:2012aa,ATLAS:2014ipv,CMS:2016uxf}.}
        \label{fig:had_and_jet_RAA}}
        \end{center}
        \end{minipage}
    \end{figure}
\vspace{3mm}

\emph{Results} --- 
Our calculations were performed using the \JETSCAPE\, framework simulating a \PbPb\, collision at $\sqrt{s}=2.76$ ATeV. Two sets of simulations were made using \CUJET\, and \Martini\, used as the low-virtuality energy loss modules with all other parameters held fixed. \Pythia\, was used to generate the hard scattering event, performing the initial state shower as well as handling multi-parton interactions. High virtuality partons were then handed over to \Matter\cite{Majumder:2013re,Cao:2017qpx}, to be further evolved down in virtuality.  
For the \pp\, spectra used in the \RAA\, calculation and the tuning of the other modules in \JETSCAPE\, Ref$.$~\cite{JETSCAPE:2019udz} was used.
The hydrodynamic background was provided by a \Vishnu\, $\left(2+1\right)$D viscous hydrodynamic simulation 
with temperature dependent specific shear and bulk viscosities~\cite{Bernhard:2019bmu}. Finally \FastJet~\cite{Cacciari:2011ma} was used for the jet-clustering.

\begin{figure}[!hbt]
        \includegraphics[width=0.7\linewidth]{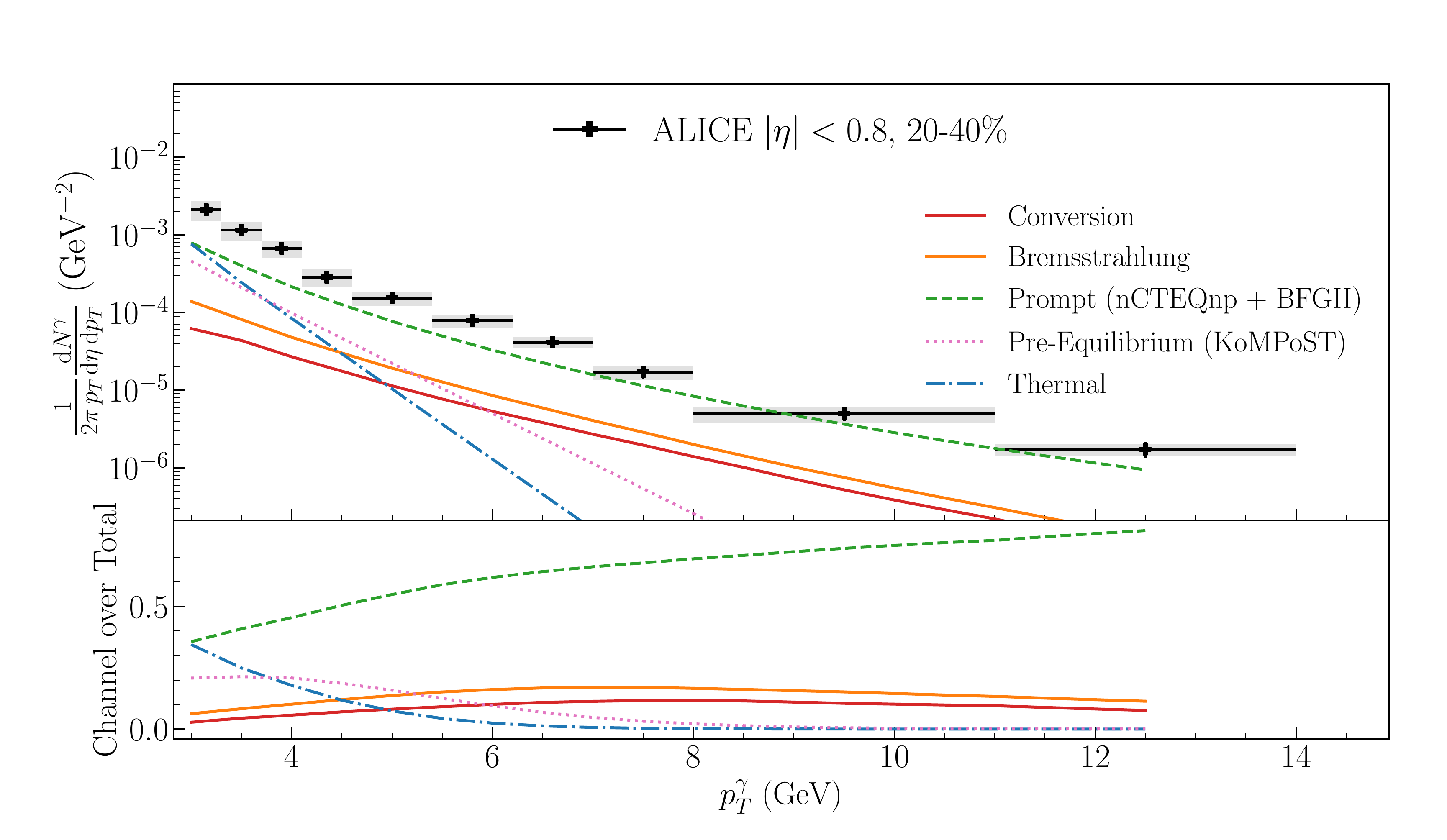}
        \begin{minipage}[b]{6pc}
        \begin{center} 
        \caption{Channel break down of the direct photon yield at midrapidity of $20$-$40\%$ collisions of \PbPb\, at $\sqrt{s}=2.76$ ATeV. Data from \protect~\cite{ALICE:2015xmh}.
        \label{fig:channnel_by_channel_photons}}
        \end{center}
        \end{minipage}
    \end{figure}

Fig$.$~\ref{fig:had_and_jet_RAA} (a)  and (b) show the results of the simulation for charged hadrons and jets, respectively, for the $0$-$5\%$ centrality class. We see satisfactory agreement between the two energy
loss modules and the data. Despite the theoretical differences between them, the two energy loss models are nearly indistinguishable. We now turn our attention to photons. Fig$.$~\ref{fig:channnel_by_channel_photons} provides a channel-by-channel calculation of photon yield at mid-rapidity for the same colliding system as before (at $20$-$40\%$ centrality) and compared to data from ALICE. The prompt, pre-equilibrium and thermal photons are taken from Ref$.$~\cite{Gale:2021emg}.

\begin{figure}[!hbt]
        \includegraphics[width=0.6\linewidth]{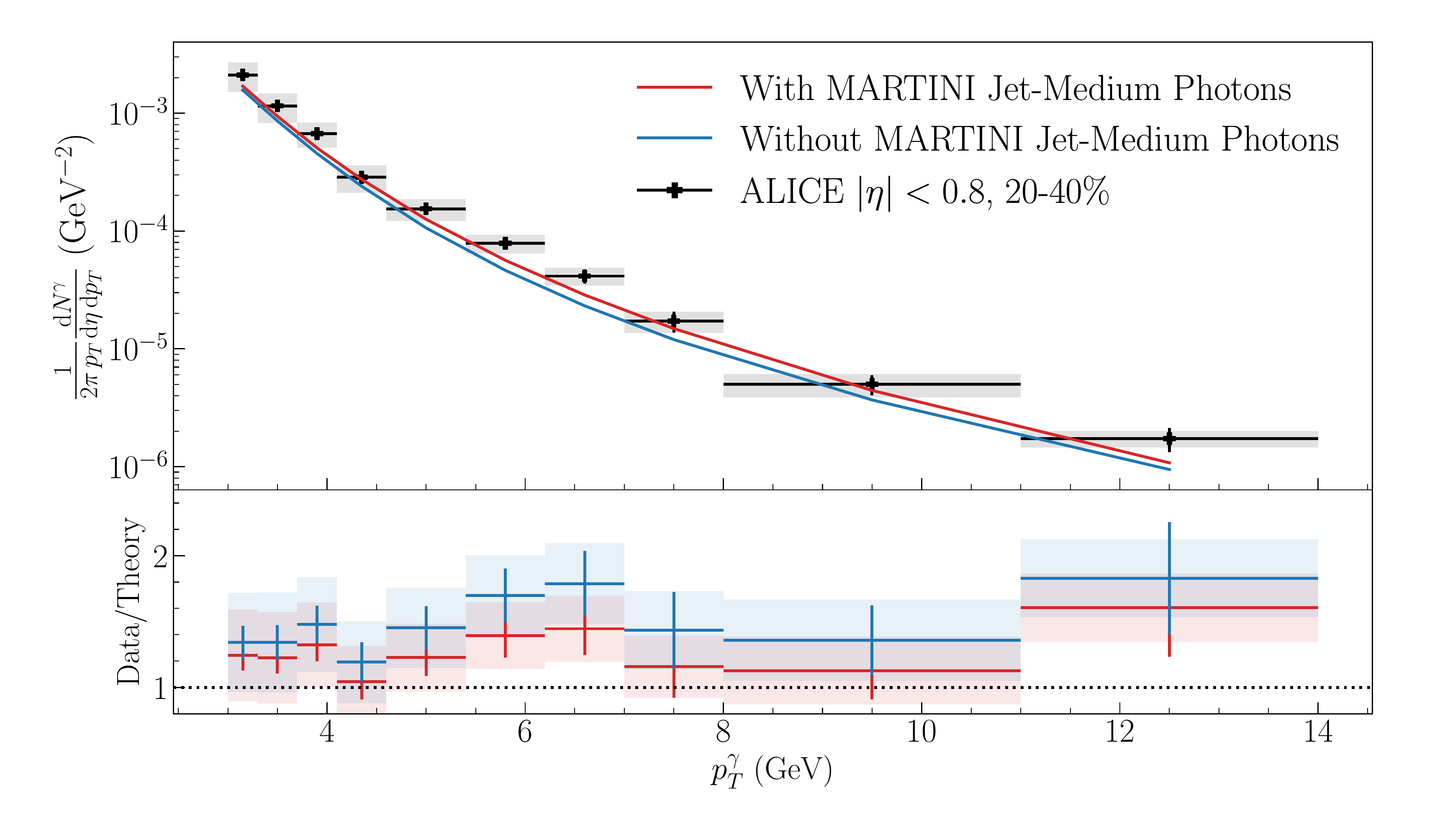}
        \begin{minipage}[b]{11pc}
        \begin{center} 
        \caption{
        Transverse momentum spectrum of direct photons, with and without jet-medium contribution. Jet-medium photons are calculated using \Martini. Error bars and shaded regions are statistical and systematic uncertainties, respectively.\label{fig:with_and_without_jet_medium}}
        \end{center}
        \end{minipage}
    \end{figure}

It is clear from Fig$.$~\ref{fig:channnel_by_channel_photons} that the significance of medium sources of photons (pre-equilibrium and thermal) goes down while that of the prompt photons goes up as we consider larger and larger values of photon $p_T$. The jet-medium channels contribute $\approx 30\%$ at intermediate values of photon $p_T$, peaking at $\approx 7$ GeV. Fig$.$~\ref{fig:with_and_without_jet_medium} makes the effect of including jet-medium photons even more stark. Including photons from jet-medium interactions brings the overall curve into much better 
agreement with the data, particularly for the intermediate $p^{\gamma}_T\in \left[5,8\right]$ GeV.

\begin{figure}[!hbt]
        \includegraphics[width=0.65\linewidth]{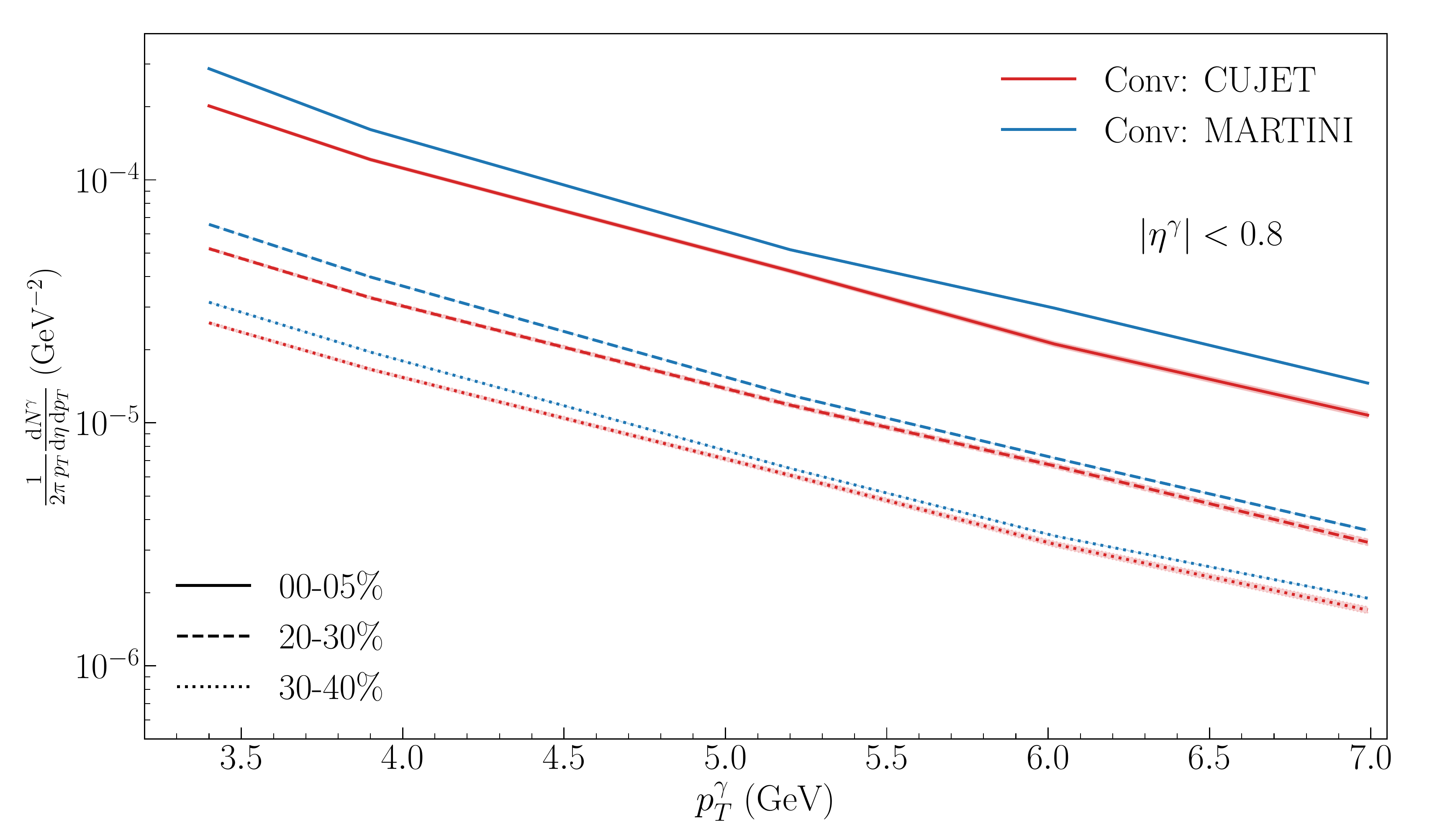}
    \begin{minipage}[b]{10pc}
    \begin{center} 
        \caption{ Comparison of conversion photon spectra from \CUJET\, and \Martini, for 3 centrality classes of \PbPb\, at $\sqrt{s}=2.76$ ATeV. See text for details.
        \label{fig:compare_CUJET_MARTINI}}
    \end{center}
    \end{minipage}
    \end{figure}

To compare the equivalent photon yield from \CUJET\, we require a \CUJET\, calculation of bremsstrahlung photons , which is currently being completed.
Thus we restrict the comparison to conversion photons in Fig$.$~\ref{fig:compare_CUJET_MARTINI} 
for three centrality classes. The more peripheral collisions result in a smaller medium and lower
temperatures. Given the temperature dependence of the rates of conversion photons, it is natural to
see the $30$-$40\%$ curves to be nearly identical. The $0$-$5\%$ centrality class shows larger difference
between the two modules: conversion photon yield from \Martini\, is nearly $40\%$ larger than that of \CUJET. The direct proportionality of conversion photons to the underlying
$q,\bar{q}$ distribution, then, indicates a difference in the two modules and how they modify the evolving $q/\bar{q}$ distribution.

\vspace{3mm}
\emph{Conclusion \& Outlook} --- 
Jet energy loss and suppression of charged hadron yield relative to the proton-proton baseline is an 
important signal of the creation of QGP. In this work we presented the results of the implementation
of \CUJET\, into \JETSCAPE\, and a comparison of charged hadron and jet \RAA\, from \CUJET\, and
\Martini. We further presented the first dynamic calculation of jet-medium photons in a realistic
plasma and a dynamic initial jet distribution with \Martini. The inclusion of jet-medium photons in
Fig$.$~\ref{fig:with_and_without_jet_medium} made a clear contribution to the total photon yield,
bringing the theoretical curve into  better agreement with the experimental measurement. We saw
that the two jet-medium channels, bremsstrahlung and conversion photons, together contribute nearly
$\approx 30\%$ to the total photon yield for intermediate values of photon transverse momentum. The
comparison of \CUJET\, and \Martini\, in photon yield is currently limited to conversion photons only, since the former does not contain bremsstrahlung photons. We found that the difference in conversion photon yield from the two modules can be significant: $40\%$ in central ($0$-$5\%$) collisions which further motivates the usage of jet-medium photons as  clean probes of energy loss models.
\bibliographystyle{unsrt}
\bibliography{references}

\begin{thebibliography}{10}

\bibitem{JETSCAPE:2019udz}
A.~Kumar et~al.
\newblock {JETSCAPE framework: $p+p$ results}.
\newblock {\em Phys. Rev. C}, 102(5):054906, 2020.

\bibitem{Xu:2014ica}
Jiechen Xu et~al.
\newblock {Azimuthal jet flavor tomography with CUJET2.0 of nuclear collisions
  at RHIC and LHC}.
\newblock {\em JHEP}, 08:063, 2014.

\bibitem{Schenke:2009gb}
Bjoern Schenke et~al.
\newblock {MARTINI: An Event generator for relativistic heavy-ion collisions}.
\newblock {\em Phys. Rev. C}, 80:054913, 2009.

\bibitem{Gyulassy:2000er}
M.~Gyulassy et~al.
\newblock {Reaction operator approach to non-Abelian energy loss}.
\newblock {\em Nucl. Phys. B}, 594:371--419, 2001.

\bibitem{Djordjevic:2003zk}
Magdalena Djordjevic et~al.
\newblock {Heavy quark radiative energy loss in QCD matter}.
\newblock {\em Nucl. Phys. A}, 733:265--298, 2004.

\bibitem{THOMA1991491}
Markus~H. Thoma et~al.
\newblock Quark damping and energy loss in the high temperature qcd.
\newblock {\em Nuclear Physics B}, 351(3):491--506, 1991.

\bibitem{Fries:2002kt}
Rainer~J. Fries et~al.
\newblock {High-energy photons from passage of jets through quark gluon
  plasma}.
\newblock {\em Phys. Rev. Lett.}, 90:132301, 2003.

\bibitem{Arnold:2001ba}
Peter~B. Arnold et~al.
\newblock {Photon emission from ultrarelativistic plasmas}.
\newblock {\em JHEP}, 11:057, 2001.

\bibitem{ALICE:2012aqc}
Betty Abelev et~al.
\newblock {Centrality Dependence of Charged Particle Production at Large
  Transverse Momentum in Pb--Pb Collisions at $\sqrt{s_{\rm{NN}}} = 2.76$ TeV}.
\newblock {\em Phys. Lett. B}, 720:52--62, 2013.

\bibitem{ATLAS:2015qmb}
Georges Aad et~al.
\newblock {Measurement of charged-particle spectra in Pb+Pb collisions at
  $\sqrt{{s}_\mathsf{{NN}}} = 2.76$ TeV with the ATLAS detector at the LHC}.
\newblock {\em JHEP}, 09:050, 2015.

\bibitem{CMS:2012aa}
Serguei Chatrchyan et~al.
\newblock {Study of high-pT charged particle suppression in PbPb compared to
  $pp$ collisions at $\sqrt{s_{NN}}=2.76$ TeV}.
\newblock {\em Eur. Phys. J. C}, 72:1945, 2012.

\bibitem{ATLAS:2014ipv}
Georges Aad et~al.
\newblock {Measurements of the Nuclear Modification Factor for Jets in Pb+Pb
  Collisions at $\sqrt{s_{\mathrm{NN}}}=2.76$ TeV with the ATLAS Detector}.
\newblock {\em Phys. Rev. Lett.}, 114(7):072302, 2015.

\bibitem{CMS:2016uxf}
Vardan Khachatryan et~al.
\newblock {Measurement of inclusive jet cross sections in $pp$ and PbPb
  collisions at $\sqrt{s_{NN}}=$ 2.76 TeV}.
\newblock {\em Phys. Rev. C}, 96(1):015202, 2017.

\bibitem{Majumder:2013re}
Abhijit Majumder.
\newblock {Incorporating Space-Time Within Medium-Modified Jet Event
  Generators}.
\newblock {\em Phys. Rev. C}, 88:014909, 2013.

\bibitem{Cao:2017qpx}
Shanshan Cao et~al.
\newblock {Nuclear modification of leading hadrons and jets within a virtuality
  ordered parton shower}.
\newblock {\em Phys. Rev. C}, 101(2):024903, 2020.

\bibitem{Bernhard:2019bmu}
Jonah~E. Bernhard et~al.
\newblock {Bayesian estimation of the specific shear and bulk viscosity of
  quark\textendash{}gluon plasma}.
\newblock {\em Nature Phys.}, 15(11):1113--1117, 2019.

\bibitem{Cacciari:2011ma}
Matteo Cacciari et~al.
\newblock {FastJet User Manual}.
\newblock {\em Eur. Phys. J. C}, 72:1896, 2012.

\bibitem{Gale:2021emg}
Charles Gale et~al.
\newblock {Multimessenger heavy-ion collision physics}.
\newblock {\em Phys. Rev. C}, 105(1):014909, 2022.

\bibitem{ALICE:2015xmh}
Jaroslav Adam et~al.
\newblock {Direct photon production in Pb-Pb collisions at $\sqrt{s_{NN}} =$
  2.76 TeV}.
\newblock {\em Phys. Lett. B}, 754:235--248, 2016.

\end{thebibliography}
\end{document}